\documentclass{aa}  

\usepackage{graphicx}
\usepackage{txfonts}
\usepackage{color}
\usepackage{multirow}

\newcommand{\tablenotea}[1]{\parbox{12.1cm}{\indent \footnotesize{#1}}}

\newcommand{\chemrev}{Chem. Rev.}

\newcommand{\jms}{J. Mol. Spectr.}

\newcommand{\jpcrd}{J. Phys. Chem. Ref. Data}

\begin{document}

\title{A new protonated molecule discovered in \mbox{TMC-1}: HCCNCH$^+$\thanks{Based on observations carried out with the Yebes 40m telescope (projects 19A003, 20A014, 20D023, and 21A011) and the IRAM 30m telescope. The 40m radio telescope at Yebes Observatory is operated by the Spanish Geographic Institute (IGN; Ministerio de Transportes, Movilidad y Agenda Urbana). IRAM is supported by INSU/CNRS (France), MPG (Germany), and IGN (Spain).}}

\titlerunning{Detection of HCCNCH$^+$ in \mbox{TMC-1}}
\authorrunning{Ag\'undez et al.}

\author{M.~Ag\'undez\inst{1}, C.~Cabezas\inst{1}, N.~Marcelino\inst{2,3}, R.~Fuentetaja\inst{1}, B.~Tercero\inst{2,3}, P.~de~Vicente\inst{3}, \and J.~Cernicharo\inst{1}}

\institute{
Instituto de F\'isica Fundamental, CSIC, Calle Serrano 123, E-28006 Madrid, Spain\\ \email{marcelino.agundez@csic.es, jose.cernicharo@csic.es} 
\and
Observatorio Astron\'omico Nacional, IGN, Calle Alfonso XII 3, E-28014 Madrid, Spain 
\and
Observatorio de Yebes, IGN, Cerro de la Palera s/n, E-19141 Yebes, Guadalajara, Spain
}

\date{Received; accepted}

 
\abstract
{In recent years we have seen an important increase in the number of protonated molecules detected in cold dense clouds. Here we report the detection in \mbox{TMC-1} of HCCNCH$^+$, the protonated form of HCCNC, which is a metastable isomer of HC$_3$N. This is the first protonated form of a metastable isomer detected in a cold dense cloud. The detection was based on observations carried out with the Yebes 40m and IRAM 30m telescopes, which revealed four harmonically related lines. We derive a rotational constant $B$\,=\,4664.431891\,$\pm$\,0.000692 MHz and a centrifugal distortion constant $D$\,=\,519.14\,$\pm$\,4.14 Hz. From a high-level ab initio screening of potential carriers we confidently assign the series of lines to the ion HCCNCH$^+$. We derive a column density of (3.0\,$\pm$\,0.5)\,$\times$\,10$^{10}$ cm$^{-2}$ for HCCNCH$^+$, which results in a HCCNCH$^+$/HCCNC abundance ratio of 0.010\,$\pm$\,0.002. This value is well reproduced by a state-of-the-art chemical model, which however is subject to important uncertainties regarding the chemistry of HCCNCH$^+$. The observational and theoretical status of protonated molecules in cold dense clouds indicate that there exists a global trend in which protonated-to-neutral abundance ratios MH$^+$/M increase with increasing proton affinity of the neutral M, although if one restricts to species M with high proton affinities ($>$\,700 kJ mol$^{-1}$), MH$^+$/M ratios fall in the range 10$^{-3}$-10$^{-1}$, with no apparent correlation with proton affinity. We suggest various protonated molecules that are good candidates for detection in cold dense clouds in the near future.}


\maketitle

\section{Introduction}

Ion-neutral reactions, together with neutral-neutral reactions, are thought to be responsible for the chemical synthesis of molecules in cold dense clouds \citep{Agundez2013}. Ions are therefore key intermediates in the built-up of chemical complexity in these environments. However, among the nearly 130 molecules detected toward \mbox{TMC-1}, the vast majority are neutral species, with only 11\,\% being positive ions and 4\,\% negative ions. Detecting ions has turned out to be a more difficult task than observing neutral species, most likely due to their lower abundances compared to neutrals. However, since many chemical routes involve positive ions, detecting them and constraining their abundances is essential to validate the chemical networks used to model cold dense clouds.

Positive ions detected toward cold dense clouds comprise, in addition to the widespread HCO$^+$ and N$_2$H$^+$, HCS$^+$ \citep{Thaddeus1981}, HCNH$^+$ \citep{Schilke1991}, HC$_3$NH$^+$ \citep{Kawaguchi1994}, HCO$_2^+$ \citep{Turner1999}, NH$_3$D$^+$ \citep{Cernicharo2013}, NCCNH$^+$ \citep{Agundez2015}, H$_2$COH$^+$ \citep{Bacmann2016}, NS$^+$ \citep{Cernicharo2018}, and H$_2$NCO$^+$ \citep{Marcelino2018}. In the last two years, the list has increased considerably with the discovery of the cations HC$_5$NH$^+$ \citep{Marcelino2020}, HC$_3$O$^+$, HC$_3$S$^+$, CH$_3$CO$^+$, C$_3$H$^+$, C$_5$H$^+$ \citep{Cernicharo2020a,Cernicharo2021a,Cernicharo2021b,Cernicharo2022}, HCCS$^+$ and HC$_7$NH$^+$  \citep{Cabezas2022a,Cabezas2022b}. Most of the cations observed, in fact all except NS$^+$, are the protonated form of abundant neutral molecules.

The chemistry of protonated molecules in cold dense clouds has been described by \cite{Agundez2015}. In a simplified chemical scheme, protonated molecules are mostly formed by proton transfer to the neutral counterpart from abundant proton donors, although ion-neutral reactions different to proton transfer may compete, and they are mainly destroyed by dissociative recombination with electrons. Protonated molecules have abundances in the range 0.01-10 \% of their corresponding neutral counterpart. In fact, there exists a trend in which the protonated-to-neutral abundance ratio increases with increasing proton affinity of the neutral \citep{Agundez2015}. Although the chemical composition of cold dense clouds is known to be regulated by chemical kinetics, rather than thermochemical considerations, the proton affinity matters in establishing how abundant a protonated species can be, most probably because a high proton affinity increases the number of possible proton donors and thus the overall protonation rate. It is worth noting that chemical models of cold dense clouds tend to underestimate protonated-to-neutral abundance ratios, which indicates that chemical networks are probably not accurate enough yet.

Here we report the discovery of a new protonated molecule in the cold dense cloud \mbox{TMC-1}, HCCNCH$^+$. This ion is the protonated form of HCCNC, a metastable isomer of HC$_3$N, which is relatively abundant in \mbox{TMC-1}. We also review briefly the situation concerning protonated molecules in cold dense clouds, and discuss the prospects for finding new ions of this type.

\section{Observations}

The observational data used in this article consist of spectra of \mbox{TMC-1} taken with the Yebes 40m and IRAM 30m telescopes. The observed position corresponds to the cyanopolyyne peak of \mbox{TMC-1}, $\alpha_{J2000}=4^{\rm h} 41^{\rm  m} 41.9^{\rm s}$ and $\delta_{J2000}=+25^\circ 41' 27.0''$. 

The Yebes 40m observations are part of the ongoing line survey of \mbox{TMC-1} QUIJOTE\footnote{Q-band Ultrasensitive Inspection Journey to the Obscure TMC-1 Environment.} \citep{Cernicharo2021c}. Observations were carried out during different observing runs between November 2019 and May 2021. The total on-source telescope time is 238 h in each polarization (twice this value after averaging the two polarizations). The QUIJOTE line survey uses a 7 mm receiver covering the Q band (31.0-50.3 GHz) with horizontal and vertical polarizations. Receiver temperatures during 2019 and 2020 varied from 22 K at 32 GHz to 42 K at 50 GHz. In 2021, some power adaptation carried out in the down-conversion chains lowered the receiver temperatures to 16\,K at 32 GHz and 25\,K at 50 GHz. The backends are fast Fourier transform spectrometers which provide a bandwidth of 8\,$\times$\,2.5 GHz in each polarization, thus covering practically the whole Q band, with a spectral resolution of 38.15 kHz. The system is described in detail by \citet{Tercero2021}. The QUIJOTE observations were performed using the frequency-switching observing mode with frequency throws of 8 or 10 MHz, depending on the observing session. The half power beam width (HPBW) of the Yebes 40m telescope in the Q band can be fitted as a function of frequency as HPBW($''$)\,=\,1764/$\nu$(GHz).
 
The IRAM 30m data correspond to observations in the 3\,mm band and include data from our 3\,mm line survey of \mbox{TMC-1} \citep{Marcelino2007,Cernicharo2012} and data taken in September 2021, which are described in \cite{Agundez2022} and \cite{Cabezas2022a}. Briefly, in the observations carried out in 2021 the 3\,mm EMIR receiver was used connected to a fast Fourier transform spectrometer, providing a spectral resolution of 48.84 kHz. We used two frequency setups at slightly different central frequencies to check for spurious signals, image band contamination, and other technical artifacts. The frequency-switching observing mode was used with a frequency throw of 18\,MHz. After averaging all IRAM 30m data, the total on-source integration time is 35.4 h for each polarization. For the IRAM 30m telescope, HPBW($''$)\,=\,2460/$\nu$(GHz)

The intensity scale in the Yebes 40m and IRAM 30m telescopes is antenna temperature, $T_A^*$, which has an estimated uncertainty of 10~\% and can be converted to main beam brightness temperature, $T_{mb}$, by dividing by $B_{\rm eff}$/$F_{\rm eff}$. For the Yebes 40m telescope\footnote{\texttt{https://rt40m.oan.es/rt40m\_en.php}}, $B_{\rm eff}$ can be fitted as a function of frequency as $B_{\rm eff}$\,=\,0.738\,$\exp$[$-$($\nu$(GHz)/72.2)$^2$] and $F_{\rm eff}$\,=\,0.97. For the IRAM 30m telescope\footnote{\texttt{https://publicwiki.iram.es/Iram30mEfficiencies}}, $B_{\rm eff}$\,=\,0.871\,$\exp$[$-$($\nu$(GHz)/359)$^2$], and $F_{\rm eff}$\,=\,0.95 in the 3\,mm band. All data were analyzed using the GILDAS software\footnote{\texttt{http://www.iram.fr/IRAMFR/GILDAS/}}.

\section{Results}

\begin{figure}
\centering
\includegraphics[angle=0,width=\columnwidth]{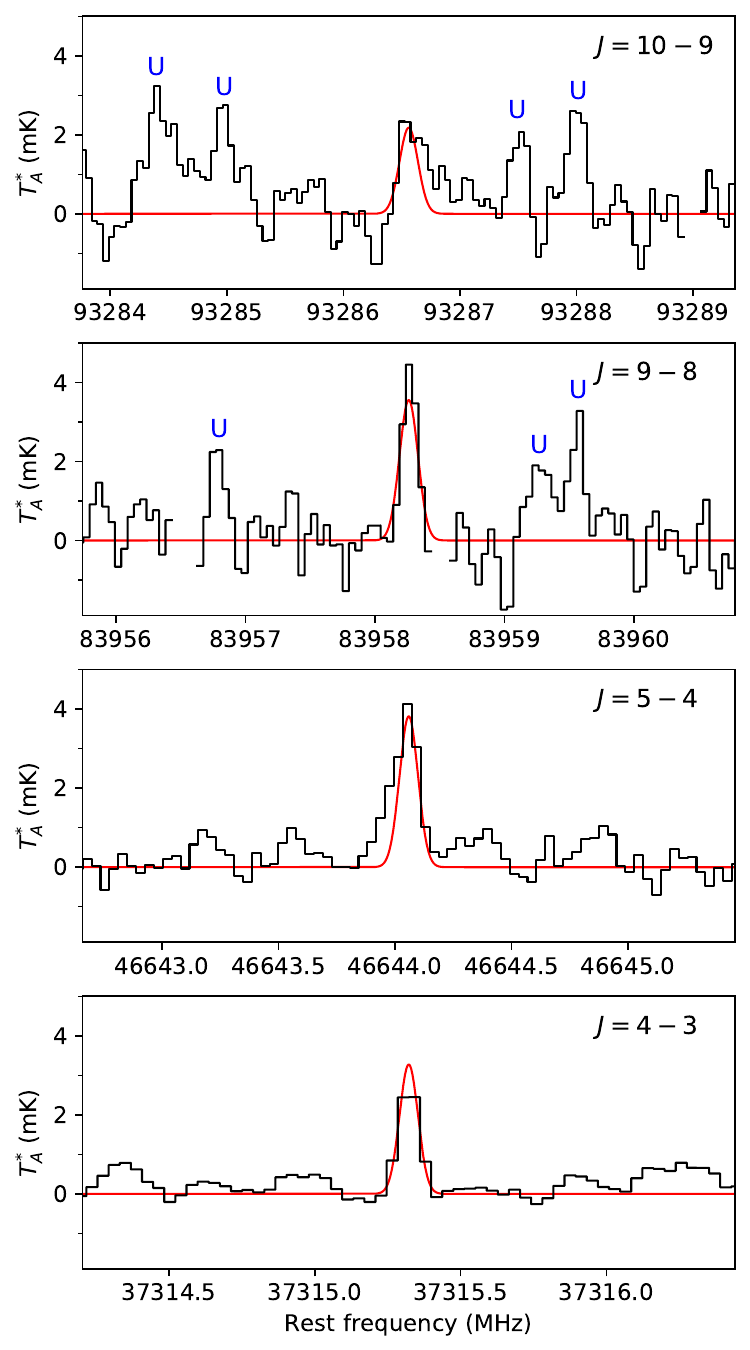}
\caption{Lines of HCCNCH$^+$ observed toward \mbox{TMC-1}. The $J$\,=\,4-3 and $J$\,=\,5-4 lines were observed with the Yebes 40m telescope and the $J$\,=\,9-8 and $J$\,=\,10-9 lines with the IRAM 30m telescope. Unidentified lines are labeled as "U". The red lines correspond to the line profiles calculated with the LVG method adopting a column density for HCCNCH$^+$ of 3.0\,$\times$\,10$^{10}$ cm$^{-2}$, a kinetic temperature of 10 K, a volume density of H$_2$ of 4\,$\times$\,10$^4$ cm$^3$, a linewidth of 0.60 km s$^{-1}$, and an emission size of 40$''$ in radius (see text).} \label{fig:lines}
\end{figure}

\begin{table*}
\small
\caption{Observed line parameters of HCCNCH$^+$ in \mbox{TMC-1}.}
\label{table:lines}
\centering
\begin{tabular}{cccccc}
\hline
\hline
\multicolumn{1}{c}{Transition} & \multicolumn{1}{c}{$\nu_{obs}\,^a$} & \multicolumn{1}{c}{$\nu_{obs}-\nu_{calc}$\,$^b$} & \multicolumn{1}{c}{$\Delta v$\,$^c$} & \multicolumn{1}{c}{$T_A^*$ peak} & \multicolumn{1}{c}{$\int T_A^* dv$} \\
          &  \multicolumn{1}{c}{(MHz)}     &    \multicolumn{1}{c}{(MHz)}               & \multicolumn{1}{c}{(km s$^{-1}$)}     &  \multicolumn{1}{c}{(mK)} &   \multicolumn{1}{c}{(mK km s$^{-1}$)}    \\
\hline
$J$\,=\,4-3  & 37315.322\,$\pm$\,0.010 & $+$0.000 & 0.66\,$\pm$\,0.03 & 2.78\,$\pm$\,0.18 & 1.96\,$\pm$\,0.11 \\
$J$\,=\,5-4  & 46644.056\,$\pm$\,0.010 & $-$0.003 & 0.68\,$\pm$\,0.08 & 4.01\,$\pm$\,0.31 & 2.92\,$\pm$\,0.29 \\
$J$\,=\,9-8  & 83958.268\,$\pm$\,0.010 & $+$0.008 & 0.46\,$\pm$\,0.07 & 4.45\,$\pm$\,0.61 & 2.18\,$\pm$\,0.31 \\
$J$\,=\,10-9 & 93286.556\,$\pm$\,0.010 & $-$0.005 & 0.59\,$\pm$\,0.13 & 2.00\,$\pm$\,0.62 & 1.25\,$\pm$\,0.41 \\
\hline
\end{tabular}
\tablenotea{\\
The line parameters $\nu_{\rm obs}$, $\Delta v$, $T_A^*$ peak, and $\int T_A^* dv$ as well as the associated errors were derived from a Gaussian fit to each line profile. The rms noise level is given as the error of $T_A^*$ peak.\\
$^a$\,Observed frequencies adopting a systemic velocity of 5.83 km s$^{-1}$ for \mbox{TMC-1} \citep{Cernicharo2020b}.\\
$^b$\,Observed minus calculated frequencies, where calculated frequencies, $\nu_{calc}$, are computed using $B$\,=\,4664.431891 MHz and $D$\,=\,519.14 Hz (see text).\\
$^c$\,$\Delta v$ is the full width at half maximum.\\
}
\end{table*}

Among the unidentified lines present in the 7\,mm and 3\,mm data of \mbox{TMC-1} there are four lines with peak antenna temperatures between 2 and 4 mK, which have a nearly perfect harmonic relation 4/5/9/10. Two lines lie in the 7\,mm band, at 37315 MHz and 46644 MHz, while the other two lie in the 3\,mm band, at 83958 MHz and 93287 MHz. The lines are shown in Fig.~\ref{fig:lines} and their accurate measured frequencies are given in Table~\ref{table:lines}. The line frequencies can be fitted to a rotational constant $B$\,=\,4664.431891\,$\pm$\,0.000692 MHz and a centrifugal distortion constant $D$\,=\,519.14\,$\pm$\,4.14 Hz, with an rms of 7 kHz. Given the nearly perfect harmonic relation, the probability that the four lines arise from different carriers is negligible. The fit predicts three additional lines in the 3\,mm band, at 74630 MHz, 102615 MHz, and 111943 MHz. However, the sensitivity of our data at these frequencies is 4.0 mK, 3.5 mK, and 4.8 mK in $T_A^*$, i.e., much poorer than for the other two lines in the 3\,mm domain.

There are various possible carriers for the series of lines observed in \mbox{TMC-1}. The derived rotational constant of 4664.4 MHz is of the order of that of HC$_3$O$^+$, 4460.5 MHz, which has been recently detected toward \mbox{TMC-1} \citep{Cernicharo2020a}. In that study, calculations at the CCSD(T)-F12/cc-pCVTZ-F12 level of theory were carried out for various species with rotational constants of the order of that of HC$_3$O$^+$. Among these species, HCCNCH$^+$ has the closest rotational constant to the one derived here. We note that there is a typographical error in the calculated value of the rotational constant of HCCNCH$^+$ given in Table~2 of \cite{Cernicharo2020a}. The correct calculated value of $B$ is 4664.74 MHz and not 4646.4 MHz, as confirmed by calculations carried out in this study at the same level of theory. The agreement with the observational value is excellent, with a difference of just 0.007\,\%. The calculated value of the centrifugal distortion constant is 458.3 Hz, which is also in good agreement with the observational value. To provide more refined values of $B$ and $D$, we use as scaling factors the experimental/calculated ratios obtained for \mbox{HCCNC}, which is adopted as reference molecule. The experimental values of $B$ and $D$ for HCCNC are 4967.838144\,$\pm$\,0.000236 MHz and 626.945\,$\pm$\,0.191 Hz, respectively \citep{Vigouroux2000}. Calculations at the same level of theory mentioned above yield $B$\,=\,4967.79 MHz and $D$\,=\,575.6 Hz for HCCNC. The experimental/calculated scaling factors are thus 1.00001 for $B$ and 1.08920 for $D$. Applying these scaling factors to the theoretical constants of HCCNCH$^+$ yield $B$\,=\,4664.79 MHz and $D$\,=\,499.2 Hz. The rotational constant of HCCNCH$^+$ is essentially unchanged after the scaling (the difference is now 0.008\,\%) because calculations match very well the experimental rotational constant of the reference molecule HCCNC. The very good agreement between the theoretical rotational constant of HCCNCH$^+$ and the astronomical one provides strong support in favor of the assignment of the series of unidentified lines observed in \mbox{TMC-1} to HCCNCH$^+$. Other plausible candidates are HCNCN$^+$, HNCNC$^+$, HCNNC$^+$, HNCCCH$^+$, and HCCCO$^-$, but their rotational constants, calculated at the same level of theory mentioned above \citep{Cernicharo2020a}, differ from the one observed in \mbox{TMC-1} by 3-11\,\%. These candidates can be ruled out because the level of calculation used provides errors smaller than 1\,\%. Other plausible candidates containing four heavy atoms (C, N, and/or O) are discarded as well (see \citealt{Cernicharo2020a}).

\begin{figure}
\centering
\includegraphics[angle=0,width=\columnwidth]{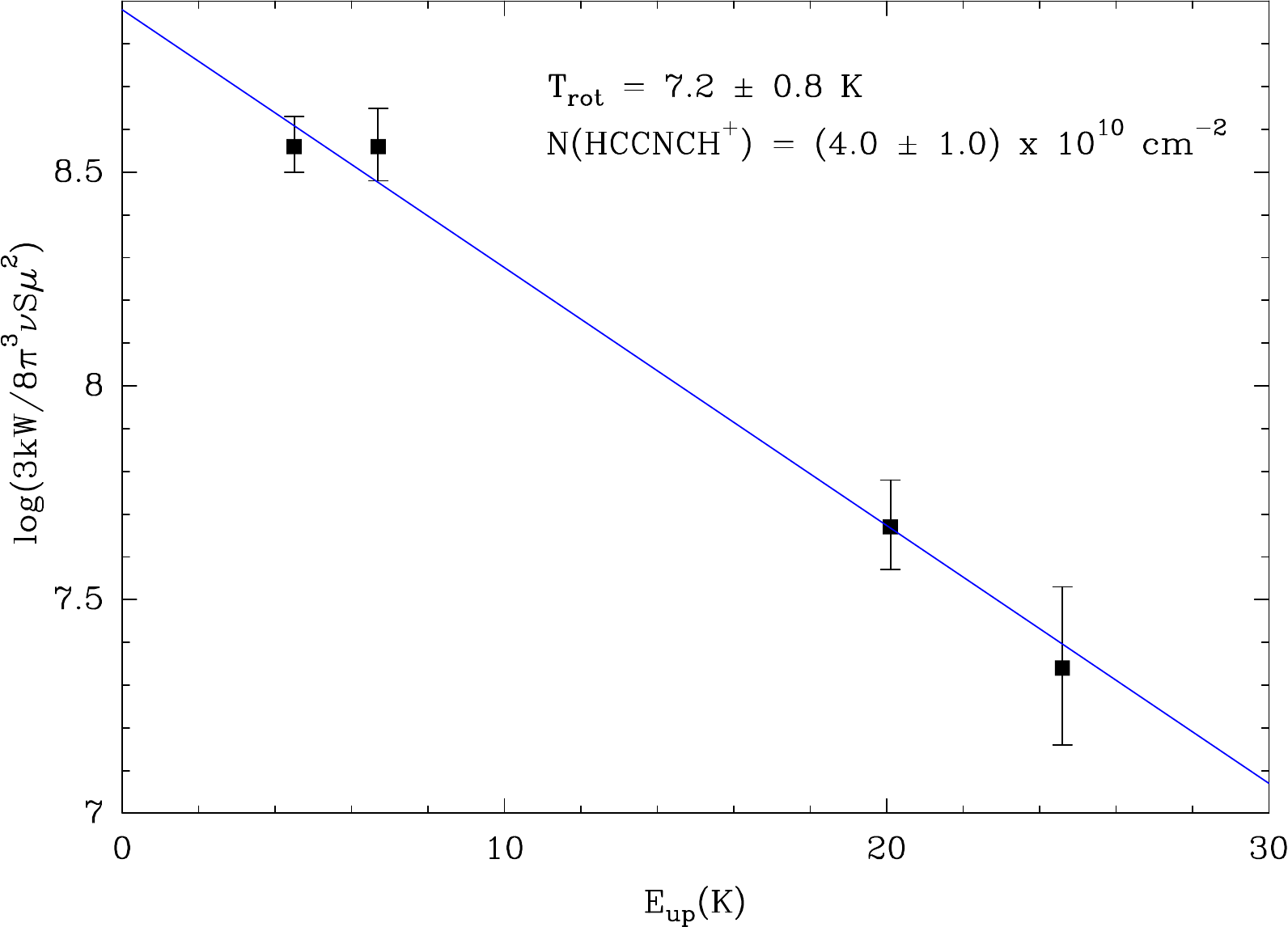}
\caption{Rotation diagram of HCCNCH$^+$ in \mbox{TMC-1}.} \label{fig:rtd}
\end{figure}

Using the line parameters derived in Table~\ref{table:lines} and a dipole moment of 3.45 Debye, as calculated for HCCNCH$^+$ by \cite{Cernicharo2020a}, we have constructed a rotation diagram, which is shown in Fig.~\ref{fig:rtd}. We derive a rotational temperature of 7.2\,$\pm$\,0.8 K, which is in the range of rotational temperatures derived in \mbox{TMC-1}, between 5 and 10 K. The column density derived from the rotation diagram is (4.0\,$\pm$\,1.0)\,$\times$\,10$^{10}$ cm$^{-2}$. We also carried out excitation calculations using the Large Velocity Gradient (LVG) method with a code written by M. Ag\'undez (see some details in \citealt{Agundez2009}), which is similar to other codes such as MADEX \citep{Cernicharo2012madex} and RADEX \citep{VanderTak2007}. We consider inelastic collisions with H$_2$ and He adopting the rate coefficients calculated for collisions between NCCNH$^+$ and He \citep{Bop2018}. We adopt an abundance of He of 0.17 relative to H$_2$ and scale the rate coefficients calculated for He, when applied to H$_2$, by multiplying them by the squared ratio of the reduced masses of the H$_2$ and He colliding systems. We adopt a gas kinetic temperature of 10 K \citep{Feher2016} and a line width of 0.60 km s$^{-1}$, the arithmetic mean of the values derived for the four lines of HCCNCH$^+$ in \mbox{TMC-1} (see Table~\ref{table:lines}). We assume that the emission is distributed in the sky as a circle with a radius of 40\,$''$, as observed for various hydrocarbons in \mbox{TMC-1} \citep{Fosse2001}. We find that the best agreement with the observed line intensities is achieved for an H$_2$ volume density of 4\,$\times$\,10$^4$ cm$^{-3}$ and a HCCNCH$^+$ column density of (3.0\,$\pm$\,0.5)\,$\times$\,10$^{10}$ cm$^{-2}$. This latter value, which is adopted hereafter, is similar to the column density derived from the rotation diagram. The calculated line profiles using these parameters are shown in Fig.~\ref{fig:lines}. The LVG calculations indicate that the $J$\,=\,4-3 and $J$\,=\,5-4 have excitation temperatures around 10 K, although the higher $J$ lines, the $J$\,=\,9-8 and $J$\,=\,10-9 have lower excitation temperatures, in the range 6-7 K. These excitation temperatures are in line with the global rotational temperature derived from the rotation diagram, 7.2 K

\begin{figure*}
\centering
\includegraphics[width=\textwidth,angle=0]{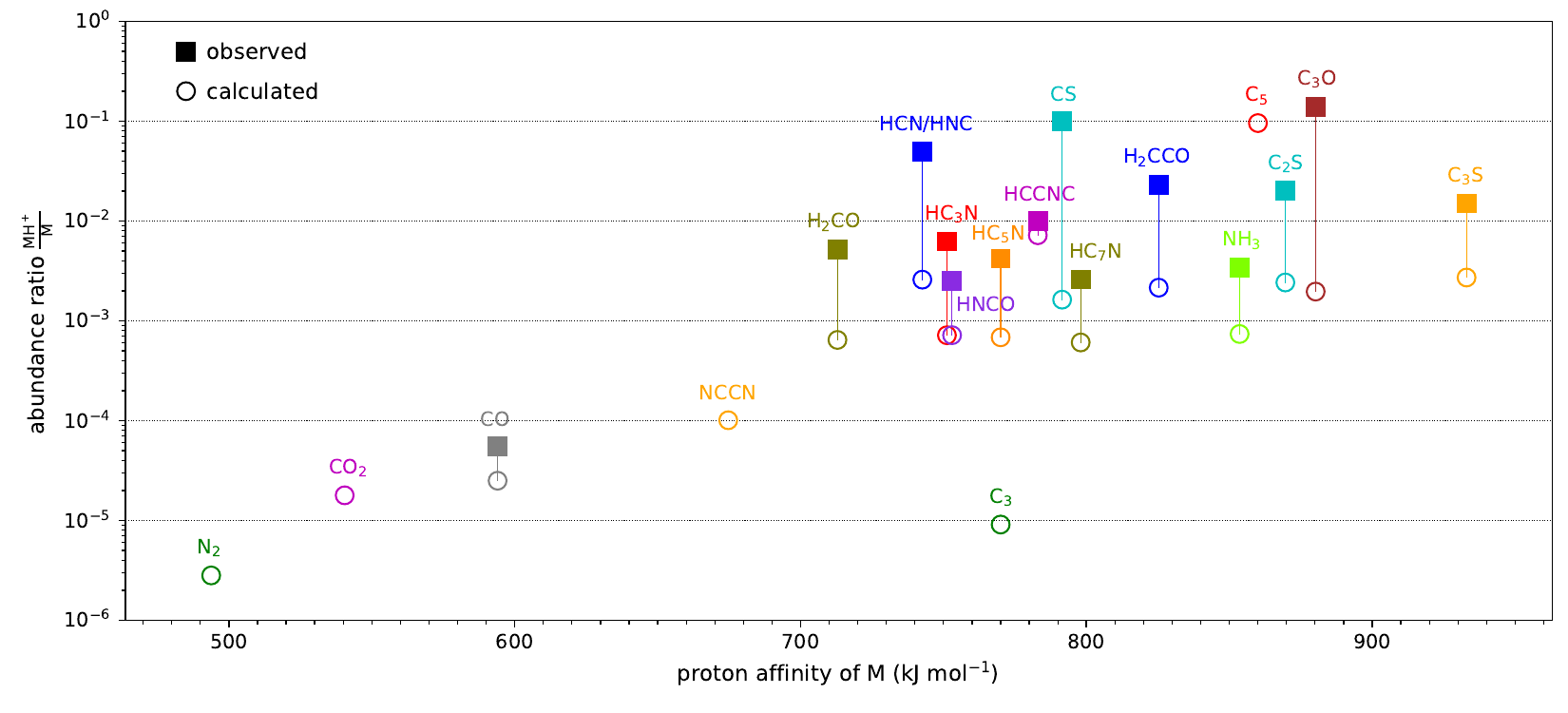}
\caption{Observed and calculated protonated-to-neutral abundance ratios, MH$^+$/M, in cold dense clouds (mostly in \mbox{TMC-1}) as a function of the proton affinity of the neutral M. The figure is based on \cite{Agundez2015}, with updates for H$_2$COH$^+$ \citep{Bacmann2016}, H$_2$NCO$^+$ \citep{Marcelino2018}, HC$_5$NH$^+$ \citep{Marcelino2020}, HC$_3$O$^+$ \citep{Cernicharo2020a}, HC$_3$S$^+$ \citep{Cernicharo2021a}, CH$_3$CO$^+$ \citep{Cernicharo2021b}, HCCS$^+$ \citep{Cabezas2022a}, HC$_7$NH$^+$ \citep{Cabezas2022b}, and HCCNCH$^+$ (this study). We also include calculated values for C$_3$H$^+$/C$_3$ and C$_5$H$^+$/C$_5$ because these two ions were recently detected toward \mbox{TMC-1} \citep{Cernicharo2022}.}
\label{fig:ratios}
\end{figure*}

\section{Discussion}

For HCCNC, \cite{Cernicharo2020b} derive a column density of (3.0\,$\pm$\,0.3)\,$\times$\,10$^{12}$ cm$^{-2}$. Therefore, the protonated-to-neutral abundance ratio HCCNCH$^+$/HCCNC is 0.010\,$\pm$\,0.002 in \mbox{TMC-1}. To put in context this result we revisit the observational and theoretical situation of protonated molecules in cold dense clouds, which is summarized in Fig.~\ref{fig:ratios}. In this figure, protonated-to-neutral abundance ratios, MH$^+$/M, are shown as a function of the proton affinity of the neutral M. Proton affinities were taken from the references listed in the caption of Fig.~\ref{fig:ratios}. To estimate the proton affinity of HCCNC we performed calculations at the CCSD/cc-pVTZ level. This method, which tends to overestimate proton affinities, yields values of 767.8 kJ mol$^{-1}$ for HC$_3$N (the experimental value is 751.2 kJ mol$^{-1}$; \citealt{Hunter1998}) and 800.1 kJ mol$^{-1}$ for HCCNC. Our best estimate of the proton affinity of HCCNC is thus 783 kJ mol$^{-1}$, which is obtained by scaling the theoretical value by the experimental/calculated ratio found for HC$_3$N.

We compiled observed MH$^+$/M ratios from the literature (see references in the caption of Fig.~\ref{fig:ratios}). The calculated MH$^+$/M values in Fig.~\ref{fig:ratios} are obtained from a pseudo time dependent gas-phase chemical model of a cold dense cloud, similar to those presented by \cite{Agundez2015}, \cite{Marcelino2020}, and \cite{Cabezas2022b}. In Fig.~\ref{fig:ratios} we plot the MH$^+$/M ratios calculated at steady state. A comparison between observed and modeled abundances indicates that the time at which chemical models best reproduce the chemical composition of \mbox{TMC-1} is in the range 10$^5$-10$^6$ yr (the so-called early-time; see e.g., \citealt{Wakelam2010,Agundez2013}). Although fractional abundances relative to H$_2$ may experience important variations between the early time and the steady state, which is usually reached at much longer times, protonated-to-neutral abundance ratios reach a steady state after some 10$^5$ yr (e.g., \citealt{Agundez2015,Cernicharo2022,Cabezas2022b}). Therefore, when dealing with protonated-to-neutral abundance ratios, as in Fig.~\ref{fig:ratios}, it is well justified to adopt calculated values at steady state. The chemical network adopted is largely based on the UMIST {\small RATE12} reaction network \citep{McElroy2013}\footnote{\texttt{http://udfa.ajmarkwick.net/}}. Among the features implemented, we updated the chemistry of HC$_7$NH$^+$ (see \citealt{Cabezas2022b}), removed the reaction between C$_2$S and O (see \citealt{Cernicharo2021a,Cabezas2022a}), and expanded the network to include the chemistry of HCCNC and its protonated form, HCCNCH$^+$, where reaction rate coefficients were mostly taken from the KIDA database \citep{Wakelam2015}\footnote{\texttt{https://kida.astrochem-tools.org/}}. 

Figure~\ref{fig:ratios} shows that there is a trend in which the protonated-to-neutral abundance ratio MH$^+$/M increases for increasing proton affinity of the neutral M. This behavior was already noticed by \cite{Agundez2015} based on a smaller number of protonated molecules detected in cold dense clouds. The rationale behind this behavior is that since proton transfer to M is one of the main formation routes to MH$^+$, the larger the proton affinity of M, the higher the number of available proton donors (all protonated species XH$^+$ such that X has a lower proton affinity than M), which results in a faster formation rate of MH$^+$ and a larger MH$^+$/M abundance ratio. However, this behavior is only noticed when looking over a broad range of proton affinities, down to that of N$_2$, and some caution must be taken because in the region of low proton affinities most species M are non polar and only calculated MH$^+$/M ratios are available. If we focus on the right side of Fig.~\ref{fig:ratios}, i.e., for proton affinities larger than 700 kJ mol$^{-1}$, then we see that there is no clear trend neither in the observed MH$^+$/M ratios nor in the calculated ones. That is, for proton affinities larger than 700 kJ mol$^{-1}$, observed and calculated MH$^+$/M ratios cluster around the region 10$^{-3}$-10$^{-1}$, with no clear correlation with the proton affinity of M. The case of C$_3$ is an exception because its protonated form, the ion C$_3$H$^+$, is the only one among the protonated molecules discovered so far in cold dense clouds that reacts rapidly with H$_2$. This fact makes C$_3$H$^+$ to reach a low abundance and thus a low C$_3$H$^+$/C$_3$ ratio \citep{Cernicharo2022}. It thus seems that there are two main regimes, depending on whether the proton affinity is lower or higher than that of CO. If the proton affinity of a molecule M is lower than that of CO, then the main proton donor to form MH$^+$ would be H$_3^+$, while if the proton affinity is higher than that of CO, then HCO$^+$, which is the most abundant ion in cold dense clouds, can act as proton donor, which enhances the formation rate of MH$^+$ by orders of magnitude. As a result, molecules M with proton affinities below that of CO reach low MH$^+$/M ratios, in the range 10$^{-6}$-10$^{-4}$, while molecules M with proton affinities higher than that of CO reach substantially higher MH$^+$/M ratios, in the range 10$^{-3}$-10$^{-1}$.
 
It is worth noting that the chemical model systematically underestimates the MH$^+$/M abundance ratio for all species considered in Fig.~\ref{fig:ratios}. To get insight into this systematic deviation we must have a look at the chemistry of protonated molecules in cold dense clouds. As discussed by \cite{Agundez2015}, in a simplified chemical scheme a protonated molecule MH$^+$ is mainly formed by proton transfer to M from the most abundant proton donors, while it is mainly destroyed through dissociative recombination with electrons. For some ions MH$^+$, ion-neutral reactions different to proton transfer may be efficient enough to become major formation routes, and this increases the MH$^+$/M over the value expected under the simplified chemical scheme in which MH$^+$ formation is dominated by proton transfer. It is likely that the lack of efficient ion-neutral reactions forming MH$^+$, other than proton transfer, in the chemical network causes the underestimation of MH$^+$/M ratios. The difference between observed and calculated MH$^+$/M ratios is most noticeable for HCS$^+$ and HC$_3$O$^+$, which suggest that there are important gaps in the chemistry of these ions in current chemical networks. The disagreement is less severe for other species such as H$_2$NCO$^+$ and HC$_7$NH$^+$.

The ion presented here, HCCNCH$^+$, shows the smallest discrepancy between calculated and observed MH$^+$/M ratio. The chemistry of HCCNCH$^+$ in the chemical network is relatively simple, but subject to important uncertainties. According to the chemical model, the main formation reaction of HCCNCH$^+$ is the ion-neutral reaction between C$^+$ and CH$_3$CN. This reaction has been measured to be rapid (the rate coefficient at 300 K is 5.6\,$\times$\,10$^{-9}$ cm$^3$ s$^{-1}$ and it is likely larger at 10 K), although information of the products is not available \citep{Anicich2003}. The KIDA database assumes that there are two channels that yield C$_2$H$_3^+$ + CN and HCCNCH$^+$ + H with equal branching ratios. Interestingly, in the chemical model this reaction is far more efficient than the proton transfer to HCCNC from abundant proton donors such as HCO$^+$ and H$_3$O$^+$, and this is behind the relatively high MH$^+$/M ratio calculated, which happens to coincide well with the observed one. Although it is uncertain whether HCCNCH$^+$ is truly produced in the reaction C$^+$ + CH$_3$CN, an interesting lesson is that for the protonated molecule with the smallest disagreement between chemical model and observations there exists an efficient ion-neutral route to the ion, different from proton transfer to the neutral counterpart. This is in line with the argument given above.

In the last years, there has been an explosion in the number of protonated molecules detected in cold dense clouds. Most of them are the protonated form of neutral species, which have an electronic closed shell and are the most stable isomer. However, recently \cite{Cabezas2022a} detected for the first time a protonated radical, HCCS$^+$, and here we report the first detection of a protonated metastable isomer, HCCNCH$^+$. It is likely that in the near future more protonated molecules will be discovered in cold dense clouds such as \mbox{TMC-1}. There are several neutral molecules which are known or expected to be abundant and whose protonated forms are good candidates for detection. Among them we can mention the carbenes H$_2$C$_3$ and H$_2$C$_4$, various non-polar or weakly polar hydrocarbons such as C$_2$H$_2$, CH$_3$CCH (and its isomer CH$_2$CCH$_2$), CH$_2$CHCH$_3$, CH$_2$CHCCH, and the cycles $c$-C$_5$H$_6$ and $c$-C$_9$H$_8$, nitriles such as CH$_3$CN, CH$_2$CHCN, and HC$_9$N, O-bearing molecules such as CH$_3$OH, CH$_3$CHO, and HC$_5$O, and S-bearing molecules such as H$_2$CS, OCS, SO, and SO$_2$. All of them are abundant in \mbox{TMC-1} and their proton affinities, when available, are above that of CO, which suggests that MH$^+$/M ratios could be in the range 10$^{-3}$-10$^{-1}$. However, for many of them, the rotational spectrum has not been characterized in the laboratory.

\section{Conclusions}

We reported the first identification in space of the ion HCCNCH$^+$ based on observations of \mbox{TMC-1} and high level ab initio calculations. We derive a HCCNCH$^+$/HCCNC abundance ratio of 0.010\,$\pm$\,0.002, which is in the range of protonated-to-neutral abundance ratios found in cold dense clouds for neutral molecules with proton affinities above 700 kJ mol$^{-1}$. A state-of-the-art chemical model reproduces well the observed HCCNCH$^+$/HCCNC ratio, although important discrepancies remain for other protonated molecules. Other neutral molecules known or expected to be abundant in cold dense clouds are likely to have relatively abundant protonated counterparts, which await detection through sensitive radioastronomical observations.

\begin{acknowledgements}

We acknowledge funding support from Spanish Ministerio de Ciencia e Innovaci\'on through grants PID2019-106110GB-I00, PID2019-107115GB-C21, and PID2019-106235GB-I00 and from the European Research Council (ERC Grant 610256: NANOCOSMOS). We thank the referee for a report that helped to clarify some points.

\end{acknowledgements}

\end{document}